%Paper: astro-ph/9406056
%From: GASPERINI@to.infn.it
%Date: Sat, 18 Jun 1994 16:40:59 +0200 (WET-DST)

\magnification=1200
\hsize 15true cm \hoffset=0.5true cm
\vsize 23true cm
\baselineskip=15pt

\font\small=cmr8 scaled \magstep0
\font\grande=cmr10 scaled \magstep4
\font\medio=cmr10 scaled \magstep2
\outer\def\beginsection#1\par{\medbreak\bigskip
      \message{#1}\leftline{\bf#1}\nobreak\medskip\vskip-\parskip
      \noindent}

\def \me {\buildrel <\over \sim}
\def \Me {\buildrel >\over \sim}
\def \pa {\partial}
\def \ra {\rightarrow}

\def \pr {\prime}
\def \se {{\prime \prime}}
\def \H {{a^\prime \over a}}

\def \ti {\tilde}

\def \Da {\Delta}
\def \b {\beta}
\def \a {\alpha}
\def \ap {\alpha^\prime}

\def \Ga {\Gamma}

\def \sg {\sigma}
\def \da {\delta}

\def \r {\rho}
\def \c {\chi}
\def \om {\omega}
\def \Om {\Omega}
\def \noi {\noindent}

\def\sqr#1#2{{\vcenter{\hrule height.#2pt\hbox{\vrule width.#2pt
height#1pt \kern#1pt\vrule width.#2pt}\hrule height.#2pt}}}

\def\lsim{\mathrel{\rlap{\lower4pt\hbox{\hskip1pt$\sim$}}
    \raise1pt\hbox{$<$}}}         %less than or approx. symbol
\def\gsim{\mathrel{\rlap{\lower4pt\hbox{\hskip1pt$\sim$}}
    \raise1pt\hbox{$>$}}}         %greater than or approx. symbol

\nopagenumbers
\line{\hfil  DFTT-24/94}
\line{\hfil June 1994}
\vskip 2 cm
\centerline {\grande  Phenomenological Aspects}
\vskip 0.5 true cm
\centerline{\grande of the Pre-Big-Bang Scenario}
\vskip 0.5 true cm
\centerline{{\grande in String Cosmology}\footnote{*}{\small
Based on talks given at {\it CERN} (June 1993), at the {\it First Int.
Workshop on Birth of the Universe and Fundamental Physics} (Rome, May
1994), and at the {\it 2nd Journ\'ee Cosmologie} (Paris, June 1994).}}

\vskip 1.5true cm
\centerline{M.Gasperini}
\centerline{\it Dipartimento di Fisica Teorica, Universit\'a di Torino,}
\centerline{\it Via P.Giuria 1, 10125 Turin, Italy,}
\centerline{\it and INFN, Sezione di Torino, Turin, Italy}

\vskip 1.5 true cm
\centerline{\medio Abstract}

\noindent
I review various aspects of the pre-big-bang scenario and of its main
open problems, with emphasis on the role played by the dilaton. Since the
dilaton is a compelling consequence of string theory, tests of this
scenario are direct tests of string theory and also, more generally, of
Planck scale physics.

\vskip 2 true cm
\centerline{---------------------------}
\centerline {To appear in the {\bf Proceedings of the 2nd Journ\'ee
Cosmologie}, }
\centerline {{\it Observatoire de Paris, June 2-4, 1994} (World Scientific
P.C., Singapore)}

\vfill\eject

\footline={\hss\rm\folio\hss}
\pageno=1

\centerline{\bf PHENOMENOLOGICAL ASPECTS OF THE PRE-BIG-BANG}
\centerline{\bf SCENARIO IN STRING COSMOLOGY}
\bigskip
\centerline{M. GASPERINI}
\centerline{\small{\it Dipartimento di Fisica Teorica, Universit\'a di
Torino, }}
\centerline{\small{\it Via P.Giuria 1, 10125 Turin, Italy,}}
\centerline{\small{\it and INFN, Sezione di Torino, Turin, Italy}}
\bigskip
\centerline{ABSTRACT}
\midinsert
\narrower
\noi
I review various aspects of the pre-big-bang scenario and of its main
open problems, with emphasis on the role played by the dilaton. Since the
dilaton is a compelling consequence of string theory, tests of this
scenario are direct tests of string theory and also, more generally, of
Planck scale physics.
\endinsert
\bigskip
\centerline{\bf Table of Contents}

1. Introduction: the pre-big-bang scenario.

2. Motivations for the pre-big-bang scenario.

3. Inflation and accelerated contraction.

4. Main open problems.

5. Phenomenological signatures of the pre-big-bang scenario:

5.1. Growing perturbation spectrum.

5.2. Cold initial state.

5.3. Dilaton production.

6. Bounds on the dilaton spectrum.

7. Conclusion.

\vskip 1 cm
\noi
{\bf 1. Introduction: the Pre-Big-Bang Scenario.}

\noi
The aim of this paper is to provide a  short but self-contained introduction
to the so-called "pre-big-bang" inflationary scenario [1], arising in the
context of string cosmology, with emphasis on its phenomenological
aspects. In particular, I will concentrate my discussion on
the process of cosmological dilaton production [2], as an example of the
fact that it is the dilaton which mainly differentiates string cosmology
from other inflationary models of the early universe.

Let me recall, first of all, that what I refer to here as "string
cosmology" is simply a model of the early universe based on the
low-energy string effective action, possibly supplemented by matter
sources. The sources may be represented, phenomenologically, even in a
perfect fluid form, but always with an equation of state which
consistently follows from the solution of the string equations of motion in
the given cosmological background. This definition is probably not enough
for a truly "stringy" description of the universe in the very high curvature
regime (see Sect. 4), but it is certainly enough for suggesting the
scenario that I have sketched in {\bf Fig. 1}.

Indeed, already at the level of the low energy string effective action,
there are motivations to expect that the present phase of standard
cosmological evolution (decelerated, with three spatial dimensions,
decreasing curvature, "frozen" Newton constant),  is preceeded in time
by a phase which is the "dual" counterpart of the present one (dual in
the sense explained in Sect. 2). Such dual phase is characterized by
accelerated expansion of the external dimensions, possible accelerated
shrinking of the internal ones, growing curvature, growing dilaton.
As a consequence, the entire time evolution of the curvature scale
corresponds to the
bell-like like curve of {\bf Fig. 1}, instead of blowing up
like in the standard model, or of approaching a constant value like in the
case of de Sitter-type evolution. A similar bell-like behaviour is expected
for the temperature and for the total effective energy density.

\vfill\eject

This cosmological picture was first sketched in some
pioneer papers based on an application of the target-space
duality of string theory in a thermodynamical context [3], and later
independently re-discovered with a different approach, based on the
solution of the string equations of motion in a curved background [4]. The
thermodynamical approach was further developed in [5,6], while
the dynamical approach led to the notion of "scale-factor" duality [7-9],
subsequently applied to formulate a possibly realistic inflationary
cosmology in [1,2,10-13].

In the context of such a scenario the "big-bang" is simply interpreted as
the phase of maximal (but finite) curvature and temperature, marking
the transition from the epoch of accelerated evolution, growing curvature
and gravitational coupling, to the standard radiation-dominated
evolution. Hence the name "pre-big-bang" for the primordial phase in
which the curvature is growing, as illustrated in {\bf Fig. 1}.

It should be
stressed, however, that {\bf Fig. 1} gives only a qualitative, very rough
description of the whole scenario. In particular, there is no need for the
evolution of the curvature scale to be time-symmetric, and a standard
inflationary phase could be included also in the post-big-bang period (or
in the transition epoch). Indeed, the pre-big-bang scenario should be
regarded not as {\it alternative}, but as {\it complementary} to the
standard (even inflationary) cosmological picture. At least
complementary to those inflationary models which cannot be extended
for ever towards the past, without running into a singularity [14].

In the following Section I will report, very briefly, some string
theory motivation supporting  a pre-big-bang cosmological
scenario.

\vskip 1 cm
\noi
{\bf 2. Motivations for the Pre-Big-Bang Scenario.}

\noi
A first motivation relies on a symmetry property of the cosmological
equations obtained from the low energy string effective Lagrangian
[15],
$$
%S=-{1\over 16\pi G_D}\intd^Dx
{\cal L}=
-\sqrt{|g|}e^{-\phi}\left[R+(\pa_\mu\phi)^2
%\pa^\mu\phi -{1\over
%12}H_{\mu\nu\a}H^{\mu\nu\a}
-{1\over 2}(\pa_{[\mu}B_{\nu\a]})^2
+ ....\right] \eqno(2.1)
$$
the so-called "scale-factor" duality [7,8]
(here $\phi$ is the dilaton and $B_{\mu\nu}=-B_{\nu\mu}$ the
antisymmetric (torsion) tensor).  This symmetry implies that if the background
fields are only
time-dependent, and if $\{a, \phi\}$ are scale factor and dilaton of a
given isotropic exact solution with $B_{\mu\nu}=0$, then a new exact
solution $\{\ti a, \ti \phi \}$ is obtained through the transformation (in
$d$ spatial dimensions)
 $$
a \ra \ti a = a^{-1}, \,\,\,\,\,\,\, \phi \ra \ti \phi = \phi -2d \ln a
\eqno(2.2 )
$$

This symmetry is a particular case of a more general
$O(d,d)$-covariance [9,16] of the cosmological equations, which holds for
spatially flat metric backgrounds, and which mixes non-trivially the
spatial components of the metric, $g_{ij}$, and of the antisymmetric
tensor, $B_{ij}$. This covariance holds even in the presence of sources,
provided they represent "bulk" string matter, satisfying the string
equations of motion in the given background [11].
In such case, the duality transformation (2.2) acting on the background
is to be accompanied (in a perfect fluid approximation) by  a reflection of
the equation of state,
$$
 p/\r \ra \ti p/\ti \r = -p/\r \eqno(2.3)
$$
It should be noted, moreover, that a generalized version of duality
symmetry can be extended also to the case of spatially curved
manifolds invariant under some non-abelian isometry [17], though there
are technical problems related to the dilaton transformation in case the
isometry groups is non-semisimple  [18].

What is important, in our context, is that by combining a duality
transformation with time-reversal, $t \ra -t$, we can
always associate to any given decelerated, expanding solution, with
decreasing curvature, characterized by the condition
$$
\dot a >0,\,\,\,\,\,\,\, \ddot a <0, \,\,\,\,\,\, \, \dot H <0 \eqno(2.4)
$$
($H=\dot a/a$, and a dot denotes differentiation with respect to the
cosmic time $t$), an accelerated, expanding solution, with growing
curvature,
$$
\dot a >0,\,\,\,\,\,\,\, \ddot a >0, \,\,\,\,\,\, \, \dot H >0 \eqno(2.5)
$$
of the pre-big-bang type [1]. Consider, as example of decelerated
"post-big-bang" solution, the simple but important case of standard
radiation-dominated evolution, with frozen dilaton,
$$
a=t^{1/2},\,\,\,\, p=\r/3,\,\,\,\, \phi= const,\,\,\,\, t>0 \eqno(2.6)
$$
which is still an exact solution of the string cosmology equations. The
associated pre-big-bang solution
$$
a=(-t)^{-1/2},\,\,\,\, p=-\r/3,\,\,\,\, \phi= -3 \ln (-t),\,\,\,\, t<0
\eqno(2.7)
 $$
describes superinflationary [19] expansion, with growing
dilaton.  Owing to the properties of the low energy string effective
action it is then possible, in particular,
to find "self-dual" cosmological  solutions [1], characterized by
the condition
$$
a(t)= a^{-1}(-t) \eqno(2.8)
$$
and connecting in a smooth way the two duality-related regimes. This is
impossible in the context of the Einstein equations, where there is no
dilaton, and this duality symmetry cannot be implemented.

A second string theory motivation for the pre-big-bang scenario is
related to the fact that when the evolution is accelerated and the
curvature scale is growing, according to eq.(2.5), then the proper size of
the event horizon
$$
d_e(t)= a(t)\int_{t}^{t_1} dt^{\pr} a^{-1}(t^{\pr}) \eqno(2.9)
$$
($t_1$ is the maximal allowed future extension of the cosmic time
coordinate in the given manifold) shrinks linearly in cosmic time,
$$
d_e(t) \sim (t_1-t) , \,\,\,\,\,\,\,\,\,\, t \ra t_1 \eqno(2.10)
$$
instead of being constant like in case of de Sitter-like exponential
expansion (see Table I and II of Ref.[1]). No problems for points,
of course, but objects of finite proper size may become in such
case larger than the horizon itself. Different points of the object, falling
along different geodesics, tend to become asymptotically causally
disconnected, because their proper spatial separation becomes larger
than the proper size of the local horizon associated to their relative
acceleration [20].

In such a situation the stress tensor  $T_\mu \,^\nu$ of
a gas of test strings satisfies the condition [4,21]
$$
T_0\,^0\simeq\sum_iT_i\,^i \eqno(2.11)
$$
which implies, in the perfect fluid approximation $T_0\,^0=\r$, $T_i\,^j =
-p\da_i\,^j$, a negative effective pressure $p<0$. With such a negative
pressure the string gas itself may sustain, self-consistently, the given
background evolution [1]. This is impossible in the case of point-like
objects, where there is no such asymptotic "stretching" regime, and then
no equation of state compatible with the background evolution.

This means, in other words, that by using as model of sources a
sufficiently diluted gas of classical strings, we can find self-consistent
solutions to the full system of equations [2,22] including {\it both} the
background field equations obtained from the Lagrangian (2.1),
$$
R_\mu\,^\nu +\nabla_\mu\nabla^\nu \phi-
%{1\over
%2} \da_\mu\,^\nu {\pa V \over \pa \phi} -
{1\over 4} H_{\mu\a
\b}H^{\nu\a\b} = 8\pi G_D e^\phi T_\mu^\nu \eqno(2.12)
$$
$$
R-(\nabla_\mu\phi)^2+2 \nabla_\mu
\nabla^\mu \phi
%+V-{\pa V\over \pa \phi}
-{1\over
12}H_{\mu\nu\a}H^{\mu\nu\a}=0\eqno(2.13)
$$
$$
\pa_\nu(\sqrt{|g|}e^{-\phi}H^{\mu\a\b})=0\eqno(2.14)
$$
with an effective source term given by a sum over all strings of the
the stress tensor of each individual strings,
$$
T^{\mu\nu}(x)= {1\over \pi \ap \sqrt{|g|}}\int d\sg d\tau (
{dX^\mu \over d\tau} {dX^\nu \over d\tau}-
{dX^\mu \over d\sg} {dX^\nu \over d\sg}) \da^D (X-x) \eqno(2.15)
$$
{\it and} the string equations of motion (plus constraints) in the same
given background,
$$
{d^2X^\mu \over d\tau^2} - {d^2X^\mu \over d\sg^2} +\Ga_{\a\b}^\mu
({dX^\a \over d\tau} + {dX^\a \over d\sg} )({dX^\b \over d\tau} -
{dX^\b \over d\sg} )=0
$$
$$
g_{\mu\nu}({dX^\mu \over d\tau} {dX^\nu \over d\tau} +
{dX^\mu \over d\sg} {dX^\nu \over d\sg} )=0,\,\,\,\,\,
g_{\mu\nu}{dX^\mu \over d\tau} {dX^\nu \over d\sg} =0 \eqno(2.16)
$$
Here  $G_D$ is the $D=(d+1)$-dimensional gravitational constant,
$(\ap)^{-1}$ is the string tension, $\Ga_{\mu\nu}^\a$  and $\nabla _\mu$
are connection and covariant derivative for the background metric
$g_{\mu\nu}$,  $H_{\mu\nu\a}= 6\pa_{[\mu} B_{\nu \a]}$ is the torsion
field strength, $X^\mu$ are the background string coordinates, and
$\tau$ and $\sg$ the usual world-sheet time and space variables (we are
using the gauge in which the world-sheet metric is conformally flat).

If we impose, as initial condition, the flat and cold string perturbative
vacuum (which is the most natural one in a pre-big-bang context, and
requires no fine-tuning), supplemented by a constant but non-vanishing
density of pressureless string matter [2],
$$
a = const,\,\,\,\,\,\,  \phi= -\infty,\,\,\,\,\,\, \r = cont, \,\,\,\,\,\, p=0
\eqno(2.17)
$$
then the general solution describes the accelerated evolution of the
weakly coupled initial regime towards a curved, dilaton-driven, strong
coupling regime, characterized by a final Kasner-like configuration,
possibly anisotropic,
$$
a_i\sim (-t)^{\b_i},\,\,\,\,\,\, \phi \sim (\sum_i \b_i -1) \ln (-t),
\,\,\,\,\,\,
 \sum_i\b_i^2=1,\,\,\,\, \,\,t<0 \eqno(2.18)
$$
in which the expanding dimensions superinflate [2].

It should be stressed, in conclusion of this Section, that the initial
condition (2.17) in not new, as it was previously reported also by the
Bible. In fact, as clearly stated in the Genesis [23],

{\it "In the beginning God created the Heaven and the Earth.

And the Earth was without form, and void;

and darkness was upon the face of the deep.

And the Breath of God

moved upon the face of the water..."}

\noi
which clearly means, in a less metaphorical language,

{\it "In the beginning God created the Background Fields and the Matter
Sources.

And the Sources were pressureless and embedded in flat space;

and this dark matter had negligible interactions ($\phi = -\infty$).

And the Dilaton

fluctuated in the string perturbative vacuum..."}

\noi
(For an explanation the next step, "And God said: {\it Let there be
light}", work is still in progress).
On account of the experience of Galilei, who got into juridical trouble for
contradicting the words of the Bible, we prefer to follow very closely the
cosmological prescriptions of the Genesis. Surprisingly enough, they
seem to be in excellent agreement with the overall picture of a
pre-big-bang scenario.

\vskip 1 cm
\noi
{\bf 3. Inflation and Accelerated Contraction.}

\noi
The pre-big-bang picture described in the previous Section was explicitly
based on the the so-called Brans-Dicke (BD) metric frame, in which the
effective Lagrangian (truncated to the metric and dilaton kinetic terms)
takes the form
$$
{\cal L}=
-\sqrt{|g|}e^{-\phi}\left[R+(\pa_\mu\phi)^2 \right]\eqno(3.1)
$$
This is the natural frame in a string theory context [24], as its metric
coincides with the sigma-model metric to which test strings are directly
coupled. In the associated Einstein (E) frame the dilaton is
minimally coupled to the metric, and the truncated Lagrangian is
diagonalized in the canonical form
$$
{\cal L}=
-\sqrt{|g|}\left[R-{1\over 2}(\pa_\mu\phi)^2 \right]\eqno(3.2)
$$

A possible difficulty for the pre-big-bang scenario follows from the
observation that any isotropic superinflationary solution, describing in
the BD frame growing curvature and {\it accelerated expansion},
$$
\dot a >0,\,\,\,\,\,\,\, \ddot a >0, \,\,\,\,\,\, \, \dot H >0 \eqno(3.3)
$$
when transformed into the E frame through the conformal
($d$-dimensional) rescaling [2,13]
$$
a \ra a\, e^{-\phi/(d-1)},\,\,\,\,\,\,\,\,\,\, dt \ra dt\,
e^{-\phi/(d-1)}\eqno(3.4)
$$
becomes an isotropic solution whose curvature scale is still
growing, but describing {\it accelerated contraction},
characterized by the conditions
$$
\dot a <0,\,\,\,\,\,\,\, \ddot a <0, \,\,\,\,\,\, \, \dot H <0 \eqno(3.5)
$$

Consider indeed, as example of superinflationary background,
the isotropic version of the solution (2.18), with $\b_i =\b= -1/\sqrt d <0$.
In the E frame it is transformed into the solution
$$
a\sim (-t)^{1/d},\,\,\,\,\,\, \phi \sim -\sqrt{2(d-1)\over d} \ln (-t),
\,\,\,\,\,\, t<0 \eqno(3.6)
$$
which satisfies the conditions of eq.(3.5).

This result would seem to imply that the inflationary properties of the
background are frame-dependent.
This fact, however, is only an apparent difficulty, as it turns out that
accelerated contraction is equally good to solve the kinematical
problems of the standard model as accelerated expansion [2,13]. Consider
for instance the so-called flatness problem, in an isotropic manifold with
scale factor $a\sim|t|^\a$. This problem is solved if the ratio between
the spatial curvature term, $k/a^2$, and the other terms of the
cosmological equations decreases enough during inflation,
$$
{k \over a^2 H^2}= {k \over \dot a^2}\sim {k\over |t|^{2(\a-1)}} \ra
0\eqno(3.7)
 $$
so as to compensate its subsequent growth during the phase of standard
evolution. This condition is satisfied in three possible cases:

1) $t \ra \infty, a \sim t^\a, \a>1$, namely during a phase of power-law
inflation, which obviously includes the limiting case of standard
exponential expansion ($\aÊ\ra \infty$).

2) $t \ra 0, a \sim (-t)^\a, \a<0$, namely by what is called pole-like, or
superinflationary expansion [19], and which corresponds to a
pre-big-bang solution in the BD frame.

3) $t \ra 0, a \sim (-t)^\a, 0<\a<1$, namely by a phase of accelerated
contraction, satisfying eq.(3.5), which is just the form of the
pre-big-bang solution when transformed to the E frame.

With similar arguments one can show that also the horizon problem can
be solved by a (long enough) phase of accelerated contraction.
Consider indeed a phase of accelerated evolution and growing curvature,
with $a \sim (-t)^\a$ for $t\ra 0_-$. In that case the event horizon shrinks
linearly (irrespectively of the sign of $\dot a$ [1]), while
the proper size of a causally connected region scales like the scale
factor. The horizon problem is solved if causally connected scales are
"pushed out" of the horizon, namely if the ratio
$$
\left( proper\,\, size \,\,event\,\, horizon \over proper \,\,size\,\,
caus.\,\, conn.\,\, reg.\right) \sim {(-t)\over a(t)} \sim (-t)^{1-\a}
\eqno(3.8)
$$
shrinks during inflation. Again, this may occur for $\a<0$, superinflation,
but also for $0<\a<1$, accelerated contraction, because in this last case
the event horizon shrinks always faster than the scale factor itself.

It is important to stress that both the "horizon ratio" (3.8), and the
previous "flatness ratio" (3.7), measure the duration of the accelerated
epoch in terms of the conformal time coordinate $\eta$, defined by $
d\eta = dt/a$, and that conformal time is the same in the E and BD frame
(see eq.(3.4)). As a consequence, if superinflation is long enough to solve
the standard kinematical problems in the BD frame, then the problems
are solved also in the E frame by the phase of accelerated contraction
[2,13]. One can show, moreover, that the number of strings per unit of
string volume is diluted in the same way in both frames [2], in spite
of the contraction. All these results, together with the fact that the
scalar and tensor perturbation spectrum is also the same in both frames
[2,13], assure the frame-independence of the inflationary properties of
the pre-big-bang background.

This independence emerges as a non-trivial consequence of dilaton
dynamics, because it is the dilaton which generates the conformal
transformation connecting the two frames.
\vskip 1 cm
\noi
{\bf 4. Main Open Problems.}

\noi
The scenario so far described leaves out the possible effects of a dilaton
potential. Indeed, at low enough energy scale, the dilaton potential
$V(\phi$) can appear at a non-perturbative level only, and it is expected
to be negligible, $V(\phi )\sim \exp[- \exp(-\phi)]$. In that case,
however, the growth of the curvature and of the dilaton coupling is
unbounded, as
$$
|H_i| \ra \infty,\,\,\,\,\,\,\,\,\,\,\, e^\phi \ra \infty\,\,\,\,\,\, \eqno
(4.1)
$$
when $t\ra 0_-$ in the asymptotic solution (2.18), and the background
unavoidably run towards a singularity.

We are thus led to what is probably the main difficulty of the whole
scenario, namely the possible occurrence of a smooth transition from
growing to decreasing curvature, with associated dilaton freezing. Such
a transition should include, moreover, some non-adiabatic
process of radiation production ({\it "Let there be light"}), in order to
solve also the entropy problem of the standard model.

Up to now, only  a few examples of smooth transitions are known, with
non-vanishing antisymmetric tensor in $D=2+1$ dimensions [10], or with
a non-local, repulsive dilaton potential [1], or with $\ap$ corrections
and moduli fields contributions [25], which simulate loop corrections. An
additional example of regular background [26], which is inhomogeneous,
can be  obtained by performing $O(d,d)$ transformations starting from the
cosmological solution of Nappi and Witten [27].

All these examples, however, are not fully "realistic", for various
reasons. A truly realistic model should include {\it both} $\ap$ and loop
corrections (which become important in the high curvature, strong
coupling regime), {\it and} a non-perturbative dilaton potential, related
to supersymmetry breaking, which forces the dilaton to a minimum,
gives it a mass, and freezes the value of the Newton constant (a possible
motivated example of dilaton potential is discussed in [28]). A graceful
exit from the phase of pre-big-bang evolution seems to be impossible
without including both contributions [29].  Moreover, limiting
temperature effects should be added near the Hagedorn scale, ad this
can modify the effective equation of state [30,31]. New exact solutions
[32] confirm the importance of the antisymmetric tensor $B_{\mu\nu}$
in generating interesting dynamics, but cannot avoid curvature
singularities.

Recent progress on exact string solutions, to all orders in $\ap$, is
encouraging [33]. The problem, however, is that the explicit form of the
full corrections is not known, and difficult to find to all orders. In order
to discuss possible phenomenological aspects of the pre-big-bang
scenario we shall thus adopt here a sort of "sudden" approximation, in
which we cut off the details of the transition regime, by matching
directly the pre-big-bang-phase to the standard radiation-dominated
evolution, at some given curvature scale $H_1$.
This will affect the high-energy "tail" of our predictions, but not the
predictions for scales much smaller than the string one, provided the
transition is localized in the high-curvature region. Work is in progress
to estimate the length of the transition regime [34].

We shall assume, moreover, that
the dilaton potential is negligible in the pre-big-bang phase, and
vanishing in the post-big-bang, where a constant, massive
dilaton background is sitting exactly at the minimum of the potential,
with negligible classical oscillations around it. But I will come back on
this point later, when discussing dilaton production.

\vskip 1 cm
\noi
{\bf 5. Phenomenological Signature of the Pre-Big-Bang Scenario.}

\noi
Summarizing the previous discussion, we can say that string theory
suggests a cosmological pre-big-bang scenario which has still many
unsolved problems. What is presently lacking, in particular, is a
detailed model for the transition from the growing to the decreasing
curvature regime, and for the process of non-adiabatic radiation
production. This scenario has also various interesting aspects, however,
such as natural initial condition, natural inflation, the initial singularity
of
the standard model is eventually smoothed out, and so on.

The crucial question, however, is the following: is it possible to test
observationally this scenario and, in particular, to
distinguish it from other inflationary models of the early universe?

As far as I know, it is the dilaton which marks the main difference
between string cosmology and other cosmological models, and which
characterizes the main phenomenological aspects of the pre-big-bang
scenario. At the present state of our knowledge, such aspects are:

1) {\it Scalar and tensor perturbation spectrum
growing with frequency} (the so-called
"blue" spectrum). This property is related to the
superinflationary kinematics [35, 36,12] and, as such, is not peculiar of
string cosmology. In a pre-big-bang context, however, superinflation is
not necessarily associated to a phase of dynamical dimensional reduction
[19], but it is typically a consequence of the dilaton dynamics.

2) {\it Squeezed vacuum}, and {\underbar not} squeezed thermal vacuum,
{\it as the final state of the amplified perturbations}. This quantum
property [37] is also not peculiar of string cosmology only, but of all those
inflationary models with a cold ($T=0$) initial state [38].

3) Possible existence of  a {\it relic background of cosmic dilatons}. This
is the new effect, peculiar of string cosmology, where in addition to the
metric there is also a dilaton field. The parametric amplification of the
dilaton fluctuations leads to dilaton production [2,39], just like the
amplification of the tensor part of the metric perturbations leads to
graviton production.

\vskip 1 cm
\noi
{\bf 5.1 Growing Perturbation Spectrum.}

\noi
The first point to be stressed, when considering a growing perturbation
spectrum, is that besides the usual bounds on the spectrum obtained from
large scale physics one must include, in general, an additional constraint
imposed by the observed closure density.

Indeed, let me recall that for a background evolution of the type

\centerline{\it inflation
$\ra$ radiation-dominance $\ra$ matter-dominance}

\noi
the spectral energy density of the amplified perturbations (in units of
$\r_c=H^2/G$) can be parametrized in terms of the inflation-radiation
transition scale $H_1$ as [12,36,40]
$$
\eqalign{\Om & (\om , t) \equiv {\om \over \r_c}{d\r \over d \om}
\simeq\cr
\simeq& GH_1^2\left(H_1\over H\right)^2 \left (a_1\over a\right)^4
\left(\om \over \om_1 \right)^{n-1}, \,\,\,\,\, \om_2<\om<\om_1 \cr
\simeq& GH_1^2\left(H_1\over H\right)^2 \left (a_1\over a\right)^4
\left(\om \over \om_1 \right)^{n-1} \left(\om \over \om_2\right)^{-2},
\,\,\,\,\, \om_0<\om<\om_2  \cr}\eqno(5.1)
$$
Here $n$ is th spectral index, $\om_1= a_1 H_1/a \sim
10^{11}\sqrt{H_1/M_p}$ Hz is the maximum amplified proper frequency
($M_p$ is the Planck mass), $\om_0\sim 10^{-18}$ Hz is the minimum
amplified frequency corresponding to a mode crossing today the Hubble
radius $H_0^{-1}$, and $\om_2 =a_2 H_2/a \sim   10^2Ê\om_0$ is the
frequency corresponding to the matter-radiation transition scale
$H_2Ê\sim 10^6 H_0$. The high frequency part of the spectrum
($\om>\om_2$) is determined by the first background transition only
(from inflation to radiation), while the low frequency part is affected by
both transitions.

The (approximate) large scale isotropy [41] of the cosmic microwave
background (CMB) imposes on the perturbations the condition
$$
\Om(\om_0,t_0)\me 10^{-10} \,\,\,\, \Longrightarrow \,\,\,\,
\log_{10}\left( H_1\over M_p\right) \me {2\over 5-n}(29n-39)\eqno(5.2)
$$
The critical density bound reads
$$
\Om(t)=\int ^{\om_1}_{\om_0} {d\om\over \om}\Om (\om,t) \me 1
\eqno(5.3)
$$
and implies,  for a growing spectrum ($n>1$),
$$
H_1\me M_p,\,\,\,\,\,\,\, \Om (t_0)\me 10^{-4}\eqno(5.4)
$$

If the growth is fast enough, the perturbations are thus
mainly constrained by the critical density bound, rather than by CMB
isotropy. This is illustrated in {\bf Fig. 2}, where the spectral density of
tensor perturbations is plotted versus the proper frequency, for three
different values of the spectral index $n$. Also plotted in {\bf Fig. 2} is the
constraint obtained from pulsar-timing data [42], $\Om(10^{-8}Hz)\me
10^{-6}$,
\topinsert
\vskip 12 true cm
\endinsert
\noi
and also (very roughly) the planned sensitivity of LIGO [43], just
to stress that there are more chances to observe a cosmic graviton
background in case of growing spectral distribution.

For a very fast growth of the spectrum it becomes impossible, of course,
to explain the observed COBE anisotropy [44], which should then be
ascribed to other causes (topological defects, etc.). What is to be
remarked however is that, even if constrained
by CMB, higher inflation scales become compatible with the isotropy
bound in the case of a growing spectrum, as illustrated in {\bf Fig.3},
again for tensor perturbations.

For flat or decreasing spectra ($n\leq 1$)
one has well known maximum allowed scale $H_1\sim 10^{-5} M_p$,
while for a fast enough growing spectrum ($n\Me 1.35$) scales as high as
Planckian are allowed (not higher, otherwise the produced gravitons
would overclose the universe). This is important in a string cosmology
context, where the natural scale for the inflation-radiation transition is
indeed the string scale, $H_1\sim (\ap)^{-1/2}\sim 10^{-1} M_p$, of
nearly Planckian order, which becomes in this case compatible with the
phenomenological bounds.

\topinsert
\vskip 12 true cm
\endinsert
It should be stressed, finally, that for Planckian values of the final
inflation scale $H_1$, the "coarse graining" entropy  $\Da S$ associated to
the cosmological amplification of the perturbations [45, 46]
$$
\Da S \sim \left ( H_1 \over M_p\right)^{3/2}\times (CMB\,\,
entropy)\eqno (5.5)
$$
becomes comparable with the entropy stored in the cosmic black-body
spectrum. Moreover, in such case the (pre-big-bang) $\ra$ (post
big-bang) transition could be even responsible for the generation of the
presently observed  thermal background radiation,
according to the mechanism proposed by Parker [47]. Indeed, the high
frequency part ($\om>\om_1$) of the radiation produced in that
transition is characterized by a Planck distribution, typical of thermal
equilibrium, at a proper temperature $T_1$ which today is given by
$$
T_1(t_0)\simeq {a_1H_1\over a(t_0)}=H_1\left(H_2\over H_1\right)^{1/2}
\left(H_0\over H_2\right)^{2/3}\sim 1^oK \left(H_1 \over M_p
\right)^{1/2}
$$
For $H_1\sim M_p$ this is exactly of the same order as the observed CMB
temperature.

\vskip 1 cm
\noi
{\bf 5.2  Cold Initial State.}

\noi
In order to discuss the second phenomenological signature of the
pre-big-bang scenario, I recall that the parametric amplification of the
cosmological perturbations may be interpreted, from a quantum point of
view, as a process of pair production in an external (gravitational) field,
described by a Bogoliubov transformation connecting the {\it in} and {\it
out} solutions of the linearized perturbation equation [40,48],
$$
\psi_k^{\prime \prime}+ \left[k^2-V(\eta)\right]\psi_k=0\eqno(5.6)
$$
Here  a prime denotes differentiation with respect to conformal time
$\eta$, $\psi_k$ is the Fourier component of the perturbation, for a mode
of comoving frequency $k=a\om$, and $V(\eta)$ is the effective potential
barrier associated to the transition of the background from accelerated to
decelerated evolution.  Such a  transformation relates the {\it in}
annihilation and creation operators, $\{b, b^\dagger\}$, to the {\it out}
operators, $\{a, a^\dagger\}$, $$
a_k=c_+(k)b_k+c_-^*(k)b_{-k}^\dagger,\,\,\,\,\,\,
a_{-k}^\dagger=c_-(k)b_k+c_+^*(k)b_{-k}^\dagger \eqno(5.7)
$$
where $c_{\pm}(k)$ are the Bogoliubov coefficients satisfying
$|c_-|^2-|c_+|^2=1$. It can be rewritten as a unitary transformation
$$
a_k = \Sigma_k b_k \Sigma _k^\dagger,\,\,\,\,\,\,\,\,\,\,
a_{-k}^\dagger = \Sigma_k b_{-k}^\dagger \Sigma _k^\dagger \eqno(5.8)
$$
generated by the "two-mode" squeezing operator [37,49],
$$
\Sigma_k =\exp (z_k^*b_kb_{-k} - z_k b_k^\dagger
b_{-k}^\dagger)\eqno(5.9)
$$
where the complex number $z$ parametrizes the Bogoliubov coefficients
as
$$
z_k=r_ke^{2i\theta_k},\,\,\,\,\,\,\,\, c_+= \cosh r_k,\,\,\,\,\,
c_-= e^{2i\theta_k} \sinh r_k \eqno(5.10)
$$
and depends thus on the dynamics of the external gravitational field
leading to the process of pair creation.

The squeezing operator describes the evolution of the initial state of the
fluctuations into a final "squeezed" quantum state. If we start from
the vacuum $|0\rangle$ we obtain, for each mode, a final squeezed
vacuum state [37], $|z_k\rangle =\Sigma_k |0\rangle$, with final number
of particles $N_k$ determined by the squeezing parameter $r=|z|$ or,
equivalently, by the coefficient $c_-$ which measures the content of
negative frequency modes in the {\it out} solution of the perturbation
equation,
$$
N_k= \langle z_k|b_k^\dagger b_k |z_k\rangle = \sinh^2 r_k =
\langle 0|a_k^\dagger a_k |0\rangle= |c_-(k)|^2 \eqno(5.11)
$$
If we start, however, with a non-trivial number state $|n_k\rangle$ or,
more generally, with a statistical mixture of number states [46, 50],
we obtain a squeezed statistical mixture, with final expectation number
of particles [51] $$
N= |c_-|^2(1+\overline n ) +\overline n (1+|c_-|^2) \eqno(5.12)
$$
depending on the squeezing parameter {\it and} on the initial average
number of particles $\overline n = \sum_n n p_n$, where $p_n$ are the
statistical weights of the mixture (here, and in what follows, the mode
index $k$ is to be understood, if not explicitly written).
Starting in particular from a state of thermal equilibrium, $\overline n=
\left (e^{\beta_0 \om} -1\right) ^{-1}$, the cosmological evolution leads
to a final "squeezed thermal vacuum", with expectation number of
particles given, in the large squeezing limit ($r>>1$), by
$$
N(\om) \simeq \sinh^2 r \coth \left( \b_0\om \over 2\right) \eqno(5.13)
$$

For all the inflationary models requiring, as initial condition, a state of
thermal equilibrium, the process of amplification of the fluctuations
starts {\it not} from the vacuum, but from the initial thermal bath. As a
consequence, the final state of the perturbations is a squeezed thermal
vacuum, instead of the pure squeezed vacuum, and the spectrum (5.1) is
modified as follows [38]
$$
\Om (\om,t) \simeq GH_1^2\left(H_1\over H\right)^2
\left (a_1\over a\right)^4
\left(\om \over \om_1 \right)^{n-1}
\coth \left( \b_0\om \over 2\right) \eqno(5.14)
$$
(with similar expression for the low frequency part, $\om<\om_2$). The
new parameter here is $\b_0=1/T_0$, namely the inverse of the
temperature of the initial bath, rescaled down (adiabatically) to the
present observation time $t_0$.

The evident effect of the initial temperature is that, because of
stimulated emission, particle production is enhanced at low frequency,
with respect to the spontaneous production from the vacuum. It is a sort
of "more power on larger scales" effect, due to the temperature. Such
effect is clearly illustrated in {\bf Fig. 4} for the case of a flat ($n=1$)
spectrum,  by comparing the vacuum case ($\b_0=\infty$)
\topinsert
\vskip 12 true cm
\endinsert
\noi
with two
cases at finite temperature (the rescaled initial temperature is given in
units of $\om_0^{-1}$).

As a consequence of this enhanced production, the inflation scale has to
be lowered, in order not to exceed the observed CMB anisotropy. This is
illustrated in  {\bf Fig. 5}, where the upper bound on the final inflation
scale $H_1$, obtained from the condition $\Om(\om_0)\me 10^{-10}$
applied to the modified spectrum (5.14), is plotted versus the spectral
index $n$, for different values of $\b_0$. The zero temperature case gives
the usual  bound, which reduces to $10^{-5} M_p$ for $n=1$. At
finite temperature, the maximum allowed scale is shifted to lower values.

The importance of this thermal effect depends of course on the present
value of the initial temperature $\b_0^{-1}$, and if the duration of the
inflationary phase  is very long, the effect may be completely washed out
by the inflationary super-cooling of the initial temperature. It turns out
[38], however, that the thermal corrections may be significant for those
inflationary models which predict just the "minimal" amount of inflation
required to solve the standard kinematical problems. In that case, the
effect of the thermal bath on the perturbation spectrum would represent
\topinsert
\vskip 12 true cm
\endinsert
\noi
a truly (hot) "remnant" of the pre-inflationary universe [52].

What I want to stress here is that, in principle, the experimental data
should be fitted by including a possible thermal dependence on the
spectrum.  If observed, however, such a dependence would be {\it in
contradiction} with the pre-big-bang scenario, in which the universe starts
to inflate from a flat and {\it cold} initial state, in such a way that no
thermal modification of the spectrum is predicted.
\vskip 1 cm
\noi
{\bf 5.3 Dilaton Production.}

\noi
The new effect, peculiar of string cosmology and of the pre-big-bang
scenario, is the parametric amplification of the quantum fluctuations of
the dilaton background, which leads to a process of cosmological dilaton
production [2,39].

In order to discuss this effect we need, first of all, the classical
equations determining the time evolution of the perturbations. Such
equations are obtained by perturbing the string cosmology equations
(2.12)-(2.14), and inserting the background solutions. It turns out, as
expected, that the dilaton perturbations are coupled to the scalar part of
the metric perturbations (in the linear approximation, tensor
perturbations are decoupled, and evolve independently).

Because of the frame-independence of the spectrum we can work in the
Einstein frame, where the field equations are simpler. In this frame,
however, the dilaton is coupled to the fluid sources, with coupling
functions determined by the conformal transformation connecting E and
BD frame. It is convenient, moreover, to use the gauge-invariant
Bardeen variables, and to work in the longitudinal gauge [48]. In the
case of isotropic backgrounds we have then one independent variable
$\psi$ for the scalar perturbations of the metric,
$$
ds^2=(1+2\psi) dt^2 -(1-2\psi)a^2|d \vec x|^2 \eqno(5.15)
$$
two independent variables for the matter perturbations in the perfect
fluid form,
$$
\da \r ,\,\,\,\,\,\,\,\,\,\, \da p, \,\,\,\,\,\,\,\,\,\, \da u_i \eqno (5.16)
$$
an, in addition, the dilaton perturbation
$$
\da \phi =\chi \eqno(5.17)
$$
(for the case of anisotropic backgrounds, more realistic in a pre-big-bang
scenario, work is still in progress [34]).

By working out the perturbation equations, and performing a Fourier
analysis, we end up with a system of four coupled differential equations.
Two of these equations define the velocity ($\da u_i$) and density ($\da
\r$) perturbations in terms of the background, and of the metric and
dilaton perturbations,
$$
\eqalign{ \da u^{i}_k=& \da u^{i} (a, \phi, \psi_k, \c_k) \cr
\da \r_k=& \da \r (a, \phi, \psi_k, \c_k) \cr}\eqno(5.18)
$$
The other two equations determine the coupled evolution of the metric
and dilaton perturbations, $\psi_k$ and $\c_k$, and can be written in
compact vector form as
$$
Z_k=\pmatrix {\psi_k \cr \c _k\cr},\,\,\,\,\,\,\,\,\,\,
Z^\se_k +2\H A Z_k^\pr +(k^2B+C)Z_k=0\eqno(5.19)
$$
where $A,B,C$ are $2 \times 2$, time-dependent mixing matrices,
determined by the  background and by the equation of state. For any
given background and equation of state one has then, in principle, a
solution determining the classical evolution of the dilaton
perturbations [2] (see also [53] for related work along these
lines, with the possible difference that a scalar field model of sources is
used, instead of perfect fluid matter).

The second step to be performed, in order to obtain the
quantum dilaton spectrum, is to express the perturbations in terms of
the correctly normalized variables satisfying canonical commutation
relations, and reducing the action to the canonical form [48,54] (namely the
variables determining the absolute magnitude of the vacuum
fluctuations). Such canonical variable are known for the metric-scalar
field system [55], or for the metric-fluid system [56], but not for the full
system with scalar field and fluid sources. In the full case one can try,
at present, a perturbative approach only (work is in progress along this
direction).

The problem can be avoided, however, if we assume that the duration of
the pre-big-bang inflation is much longer than the minimal duration
required to solve the standard kinematical problems. In fact, the
pre-big-bang solutions of the string cosmology equations (2.12)-(2.14)
are characterized by an integration constant, $T$, which fixes the time
scale at which the metric is no longer flat and starts to inflate, but also
fixes the time scale at which the matter contributions become negligible
with respect to the dilaton and metric kinetic energy [2],
$$
\eqalign{H^2 <<&\dot \phi ^2  \sim e^\phi \r, \,\,\,\,\,\,\,\,\,\,\, t<<T \cr
e^\phi \r <<& \dot \phi ^2  \sim H^2, \,\,\,\,\,\,\,\,\,\,\, t>>T
\cr}\eqno(5.20)
$$
So, if inflation is much longer than minimal, all the modes contributing to
the presently observed spectrum "hit" the effective potential barrier of
the perturbation equation (5.6) when the pre-big-bang is already
"dilaton-driven". Otherwise stated: all the scales inside our present
horizon crossed the horizon during the phase of dilaton dominance.

In such case we can neglect the matter contribution to the perturbation
equations, and we are left with the coupled metric and scalar field
variables only, from which we know very well how to extract the
spectrum [57]. By considering, in particular, a sudden transition between
the simplest example of pre-big-bang background, three-dimensional,
isotropic, dilaton-dominated, which in the E frame corresponds to
$$
a\sim (-\eta)^{1/2}, \,\,\,\,\,\,\,\, \phi \sim -\sqrt {12} \ln a \eqno(5.21)
$$
and the standard radiation evolution, with frozen dilaton background, we
find that the dilaton fluctuations are amplified with a growing spectral
density, $\Om_\c \sim \om^3$.

This is a very fast growth of the spectrum, but this special value of the
spectral index should not taken as particularly indicative, because it
could change substantially, while remaining growing, in a more realistic
scenario in which one includes the contribution of string matter, dilaton
potential and contracting internal dimensions [34].

\vskip 1 cm
\noi
{\bf 6. Bounds on the Dilaton Spectrum.}

\noi
For a phenomenological discussion of the consequences of a possible
dilaton production, associated to the pre-big-bang scenario,
I shall thus parametrize the spectrum in terms of a growing
spectral index $\da \geq 0$,
$$
\Om _\c (\om,t) \simeq GH_1^2\left(H_1\over H\right)^2
\left (a_1\over a\right)^4
\left(\om \over \om_1 \right)^{\da} \eqno(6.1)
$$
It turns out, as we shall see, that the phenomenological bounds are only
weakly dependent on $\da$, and they become completely
$\da$-independent for $\da > 1$.

The dilaton spectrum is constrained by various phenomenological bounds.
First of all we must require, according to eq.(5.3),
$$
H_1\me M_p\eqno(6.2)
$$
in order to avoid that the produced dilatons overclose the universe in the
radiation era, just like in the case of graviton production.

If dilatons would be massless, this would be, basically, the end of the
story. The dilatons, however, cannot be massless, because they
couple non-universally, and with at least gravitational strength, to
macroscopic matter [58] (with a possible exception which requires
however fine-tuning  [59]). This may be reconciled with the present tests
of the equivalence principle only if the dilaton range is smaller than about
1 cm, namely for a mass
$$
m\Me m_0 \equiv 10^{-4} \, eV \eqno(6.3)
$$
As a consequence we have, at low enough energy scales, also a
non-relativistic contribution to the dilaton energy density [39],
$$
\Om _\c (t) \simeq Gm^2\left(H_1\over H\right)^2
\left (a_1\over a\right)^3
\left(m \over H_1\right)^{\da -3\over 2} \eqno(6.4)
$$
due to non-relativistic modes, which start to oscillate at scales $H(t) \me
m$. Such a contribution grows in time with respect to the radiation
energy density, and in the matter-dominated era, $H<H_2$, it remains
frozen at a constant value which we must impose to be smaller than one,
in order to satisfy the critical density bound. This provides the constraint
$$
m\me \left(H_2 M_p^4 H_1^{\da -4}\right)^{1/(\da+1)}\eqno(6.5)
$$

Higher values of mass may be reconciled with present observations only
if the energy (6.4) stored in the coherent dilaton oscillations was
dissipated into radiation before the present epoch, at a decay scale
$$
H_d \simeq {m^3\over M_p^2} >H_0\eqno(6.6)
$$
The reheating associated to this decay generates an entropy [2,39]
$$
\Da S =\left(T_r\over T_d\right)^3=
\left(H_1^{4-\da}m^{\da-2} \over M_p^2\right)^{1/2}\eqno(6.7)
$$
(where $T_d$ is the decay temperature and $T_r\simeq \sqrt{M_pH_d}$ the
reheating temperature), and we are left with two phenomenological
possibilities [60].

If $m< 10^4\, GeV$ then $T_r$ is too low to allow a nucleosynthesis phase
subsequent to dilaton decay. We must impose that nucleosynthesis
occurred before, and that
$$
\Da S \me 10 \eqno(6.8)
$$
in order not to destroy all light nuclei already formed. If, on the contrary,
$m>10^4\, GeV$, and nucleosynthesis is subsequent to dilaton decay, then
the only possible constraint comes from primordial baryogenesis, and
imposes
$$
\Da S \me 10^5 \eqno(6.9)
$$
This last condition, however, could be evaded in case of low energy
baryogenesis [61].

The bounds so far considered refer to the case $m<H_1$. If $m>H_1$ then
all modes are always non-relativistic, with a total dilaton energy density
[2]
$$
\Om _\c (t) \simeq Gm^2\left(H_1\over H\right)^2
\left (a_1\over a\right)^3\eqno(6.10)
$$
and the only constraint to be imposed is
$$
m\me M_p\eqno(6.11)
$$
again to avoid over-critical density.

The bounds reported here define an allowed region in the ($m,H_1$)
parameter space. They have been already discussed in the past [62,60],
with reference to the cosmological production of massive scalar particles
with gravitational coupling strength, but always in the case of a flat $\da
=0$ spectral distribution. The resulting allowed region is illustrated in {\bf
Fig. 6}.

Such an allowed region is source of problems for the dilaton, because
realistic values of the inflation scale, say $H_1\Me 10^{-5}\, M_p$, are
only compatible with very high values of the dilaton mass. In particular,
a dilaton mass in the $TeV$ range, suggested by models of
supersymmetry breaking [63], turns out to be excluded for values of the
inflation scale suggested by the observed COBE anisotropy [64].

In the case of a growing dilaton spectrum the allowed region
"inflates" in parameter space, as illustrated in {\bf Fig. 7}, which reports
the previous bounds
computed for a linearly growing spectrum, $\da =1$.
There is a limit, however, on such a relaxation of bounds, because for
$\da >1$ the
spectral density is always
dominated by the highest frequency mode $\om_1$, even in the
non-relativistic regime,
\vfill\eject
\topinsert
\vskip 12 true cm
\endinsert
\noi
and the slope-dependence of the energy density
disappears, after integration over all modes [2,39]. For all $\da >1$ the
allowed region is thus the same as that of the linear case illustrated
here. For $0<\da <1$ the allowed region interpolates between the two
limiting cases of {\bf Fig. 6} and {\bf Fig. 7}.

There are two interesting consequences of the the allowed region of {\bf
Fig. 7}.

The first is that, for fast enough growing spectra, a dilaton mass in the
$TeV$ range becomes compatible with realistic inflation scales as
high as $10^{-5}\,M_p$.

The second is that for a "stringy" (nearly Planckian) inflation scale, the
dilaton mass may be very small or very large. In the small mass range,
$$
10^{-4}\, eV \me mÊ\me 1\, eVÊ\eqno(6.12)
$$
dilatons are not yet decayed as $m\me 100\, MeV$ (see eq.(6.6)), and
their present contribution to $\Om$ must be very near the critical one.
For the range (6.12) one has
indeed, from the density (6.4) evaluated for $\da=1$ in the matter era,
$$
10^{-4} \me \Om_\cÊ\me 1Ê\eqno(6.13)
$$
This suggests the possibility of "dilaton dark matter" [2,39], with
properties very similar to those of axion dark matter. My personal
belief is that the dilaton should be heavy, but the possibility of a
light dilaton cannot be ruled out on the grounds of the present discussion.

I should stress, however, that the bounds illustrated here refer to the
cosmological amplification of the quantum fluctuations of the dilaton
background. If there are, in addition, also classical oscillations of the
background around the minimum of the effective potential [65,60],
and/or other mechanisms of dilaton production, then other bounds are to
be superimposed to the previous ones.  Having neglected such additional
constraints, the allowed region of {\bf Fig. 7} is to be regarded as the {\it
maximal} allowed region in parameter space. Work is in progress [34] to
discuss dilaton production in a more realistic scenario in which the
post-big-bang era is not immediately dominated by radiation, but
includes a vacuum phase of dilaton dominance, dual to the phase of
dilaton-driven pre-big-bang.
\vskip 1 cm
\noi
{\bf 7. Conclusion.}

\noi
The main conclusion of this paper is that a lot of work is still needed to
clarify all the details of this cosmological scenario. However, I believe
that such work is worth to be done, for the following, very important
reason.

The dilaton is a compelling consequence of string theory.

The pre-big-bang scenario, and dilaton cosmology, are phenomenological
consequences of string theory.

Tests of this scenario are thus direct tests of string theory (and also,
more generally, of Planck scale physics).

\vskip 1 cm
\noi
{\bf Acknowledgements.}

\noi
I am grateful to R. Brandenberger, V. Mukhanov, A. Starobinski and A. A.
Tseytlin for helpful and clarifying discussions. I wish to thank also D.
Boyanovsky, T. Damour, M. Gleiser, J. Lidsey, D. Lyth and D. Wands for
their interest and for asking stimulating questions, and N. S\'anchez and
H. J. De Vega for their kind invitation to the {\it First} and {\it Second
Journ\'ee Cosmologie} (Paris, May '93, June '94). Finally, I am greatly
indebt to M. Giovannini,  J. Maharana, N. S\'anchez and G. Veneziano for a
fruitful collaboration which led to most of the results reported here.

\vskip 1 cm
\centerline{\bf References.}

\item{1.}M. Gasperini and G. Veneziano, Astropart. Phys. 1 (1993) 317

\item{2.}M. Gasperini and G. Veneziano, {\it Dilaton production in string
cosmology}, Phys. Rev. D50 (1994) (in press), Preprint CERN-TH.7178/94
(gr-qc/9403031)

\item{3.}E. Alvarez, Phys. Rev. D31 (1985) 418;

Y. Leblanc, Phys. Rev. D38 (1988) 3087;

R. Brandenberger and C. Vafa, Nucl. Phys. B31 (1989) 391;

E. Alvarez and M. A. R. Osorio, Int. J. Theor. Phys. 28 (1989) 940

\item{4.}M. Gasperini, N. S\'anchez and G. Veneziano, Nucl. Phys. B364
(1991) 365

\item{5.}A. A. Tseytlin and C. Vafa, Nucl. Phys. B372 (1992) 443;

A. A. Tseytlin, Class. Quantum Grav. 9 (1992) 979;

\item{6.}M. A. R. Osorio and M. A. Vasquez-Mozo, Preprint
CERN-TH.7085/93

\item{7.}G. Veneziano, Phys. Lett. B265 (1991) 287

\item{8.}A. A. Tseytlin, Mod. Phys. Lett. A6 (1991) 1721

\item{9.}K. A. Meissner and G. Veneziano, Phys. Lett. B267 (1991) 33;

Mod. Phys. Lett. A6 (1991) 3397

\item{10.}M. Gasperini, J. Maharana and G. Veneziano, Phys. Lett. B272
(1991) 277

\item{11.}M. Gasperini and G. Veneziano, Phys. Lett. B277 (1992) 256

\item{12.}M. Gasperini and M. Giovannini, Phys. Rev. D47 (1993) 1529

\item{13.}M Gasperini and G. Veneziano, Mod. Phys. Lett. A8 (1993) 3701

\item{14.}A. Vilenkin, Phys. Rev. D46 (1992) 2355

\item{15.}C. Lovelace, Phys. Lett. B135, (1984) 75;

E. S. Fradkin and A. A. Tseytlin, Nucl. Phys. B261, (1985) 1;

C. G. Callan, D. Friedan, E. J. Martinec and M. J. Perry, Nucl. Phys.
B262,

(1985) 593

\item{16.}A. Sen, Phys. Lett. B271 (1991) 295;

S. F. Hassan and A. Sen, Nucl. Phys. B375 (1992) 103

\item {17.}X. C. de la Ossa and F. Quevedo, Nucl. Phys. B403 (1993) 377

\item{18.}M. Gasperini, R. Ricci and G. Veneziano, Phys. Lett. B319 (1993)
438;

A. Alvarez, L. Alvarez-Gaum\'e  and Y. Lozano, Preprint CERN-TH.7204/94

(1994)

\item{19.}D. Shadev, Phys. Lett. B317 (1994) 155;

R. B. Abbott, S. M. Barr and S. D. Ellis, Phys. Rev. D30 (1994) 720;

E. W. Kolb, D. Lindley and D. Seckel, Phys. Rev. D30 (1994) 1205;

F. Lucchin and S. Matarrese, Phys. Lett. B164 (1985) 282

\item{20.}M. Gasperini, Phys. Lett. B258 (1991) 70;

Gen. Rel. Grav. 24 (1992) 219

\item{21.}M. Gasperini, N. S\'anchez and G. Veneziano, Int. J. Mod. Phys.
A6 (1991) 3853

\item{22.}M. Gasperini, M. Giovannini, K. A. Meissner and G. Veneziano, in
preparation

\item{23.}The Holy Bible, {\it Genesis}, (Oxford University Press, London)

\item{24.}N. S\'anchez and G. Veneziano, Nucl. Phys. B333 (1990) 253;

L.J. Garay and J. Garcia-Bellido, Nucl. Phys. B400 (1993) 416

\item{25.}I. Antoniadis, J. Rizos and K. Tamvakis, Nucl. Phys. B (1993)
(in press)

\item{26.}M. Gasperini, J. Maharana and G. Veneziano, Phys. Lett. B296
(1992) 51

\item{27.}C. Nappi and E. Witten, Phys. Lett. B293 (1992) 309

\item{28.}N. Kaloper and K. Olive, Astropart. Phys. 1 (1993) 185

\item{29.}R. Brustein and G. Veneziano, Preprint CERN-TH.7179/94
(February 1994)

\item{30.}M. Gleiser and J. G. Taylor, Phys. Lett. B164 (1985) 36

\item{31.}N. Turok, Phys. Rev. Lett. 60 (1987) 549

\item{32.}E. J. Copeland, A. Lahiri and D. Wands, Preprint SUSSEX-AST-94/6

\item{33.}E. Kiritsis and C. Kounnas, Preprint CERN-TH.7219/94
(hep-th/9404092)

E. Kiritsis, these Proceedings

\item{34.}M. Gasperini, M. Giovannini and G. Veneziano, in preparation

\item{35.}F. Lucchin and S. Matarrese, Ref. 19

\item{36.}M. Gasperini and M. Giovannini, Phys. Lett. B282 (1992) 36

\item{37.}L. P. Grishchuk and Y. Sidorov, Class. Quantum Grav. 6 (1989)
L161;

Phys. Rev. D42 (1990) 3413;

L. P. Grishchuk, Phys. Rev. Lett. 70 (1993) 2371

\item{38.}M. Gasperini, M. Giovannini and G. Veneziano, Phys. Rev. D48
(1993) R439

\item{39.}M. Gasperini, Phys. Lett. B327 (1994) 214

\item{40.}L. F. Abbott and M. B. Wise, Nucl. Phys. B244 (1984) 541;

L. P. Grishchuk, Sov. Phys. Usp. 31 (1988) 940;

B. Allen, Phys. Rev. D37 (1988) 2078;

V. Sahni, Phys. Rev. D42 (1990) 453;

L. P. Grishchuk and L. M. Solokhin, Phys. Rev. D43 (1991) 2566

\item{41.}G. F. Smooth et al., Astrophys. J. 396 (1992) 1

\item{42.}D. R. Stinebring et al.,  Phys. Rev. Lett. 65 (1990) 285

\item{43.}see for instance K. S. Thorne, in {\it "300 Years of Gravitation"},
ed. by S. W. Hawking and W. Israel (Cambridge Univ. Press, Cambridge,
England, 1988)

\item{44.}D. H. Lyth, these Proceedings

\item{45.}B. L. Hu and D. Pavon, Phys. Lett. B180 (1986) 329;

B. L. Hu and H. E. Kandrup, Phys. Rev. D35 (1987) 1776;

M. Gasperini and M. Giovannini, Phys. Lett. B301 (1993) 334;

R. H. Brandenberger, V. Mukhanov and T. Prokopec, Phys. Rev. Lett. 69

(1992) 3606;

Phys. Rev. D48 (1993) 2443

\item{46.}M. Gasperini and M. Giovannini, Class. Quantum Grav. 10 (1993)
L133

\item{47.}L. Parker, Nature 261 (1976) 20

\item{48.}V. Mukhanov, H. A. Feldman and R. Brandenberger, Phys. Rep.
215 (1992) 203

\item{49.}B. L. Schumaker, Phys. Rep. 135 (1986) 317

\item{50.}B. L. Hu, G. Kang and A. Matzac, Int. J. Mod. Phys. A9 (1994) 991;

B. L. Hu and A. Matzac, Phys. Rev. D (1994) (in press)

\item{51.}L. Parker, Phys. Rev. 183 (1969) 1057

\item{52.}M. Turner, Phys. Rev. D44 (1991) 3737

\item{53.}S. Mollerach and S. Matarrese, Phys. Rev. D45 (1992) 1961;

N. Deruelle, C. Gundlach and D. Langlois, Phys. Rev. D46 (1992) 5337;

T. Hirai and K. Maeda, Preprint WU-AP/32/93 (astro-ph/9404023);

D. Polarski and A. A. Starobinski, Preprint astro-ph/9404061

\item{54.}N. Deruelle, C. Gundlach and D. Polarski, Class. Quantum Grav. 9
(1992) 137

\item{55.}M. Sasaki, Prog. Theor. Phys. 76 (1986) 76;

V. F. Mukhanov, Sov. Phys. JEPT 67 (1988) 1297;

E. D. Stewart and D. H. Lyth, Phys. Lett. B302 (1993) 171

\item{56.}V. N. Lukash, Sov. Phys. JEPT 52  (1980) 807;

G. V. Chibisov and V. N. Mukhanov, Mon. Not. R. Astron. Soc. 200  (1982)

535

\item{57.}S. W. Hawking, Phys. Lett. B115  (1982) 295;

A. Linde, Phys. Lett. B116 (1982) 335;

A. A. Starobinski, Phys. Lett. B117 (1982) 175;

A. Guth and S. Y. Pi, Phys. Rev. Lett. 49 (1982) 1110;

J. M. Bardeen, P. S. Steinhardt and M. S. Turner, Phys. Rev. D28 (1983)

679

\item{58.}T. R. Taylor and G. Veneziano, Phys. Lett. B213 (1988) 459;

J. Ellis et al., Phys. Lett. B228 (1989) 264

\item{59.}T. Damour and A. M. Polyakov, Preprint IHES/P/94/1
(hep-th/9401069)

\item{60.}J. Ellis, D. V. Nanopoulos and M. Quiros, Phys. Lett. B174 (1986)
176;

J. Ellis, N. C. Tsamis and M. Voloshin, Phys. Lett. B194 (1987) 291

\item{61.}M. Gleiser, these Proceedings

\item{62.}A. S. Goncharov, A. D. Linde and M I. Vysotsky, Phys. Lett. B147
(1984) 279

\item{63.}T. Banks, D. V. Kaplan and A. Nelson, Preprint UCSD/PTH 93-26,
RU-37, (hep-ph/9308292)

\item{64.}M. White, Phys. Rev. D46 (1992) 4198;

T. Souradeep and V. Sahni, Mod. Phys. Lett. A7 (1992) 3541;

A. R. Liddle, Preprint SUSSEX-AST 93/7-2 (July 1993)

\item{65.}G. D. Coughlan et al., Phys. Lett. B131 (1983) 59;

G. German and G. G. Ross, Phys. Lett. B172 (1986) 305;

O. Bertolami, Phys. Lett. B209 (1988) 277;

B. de Carlos et al., Phys. Lett. B318 (1993) 447

\end